\documentstyle[12pt]{article}
\input epsf

\begin{document}

\begin{titlepage}

\begin{flushright}
CERN-TH/98-248\\
UM-TH 98-14\\
LAPTH 694-98\\
July 1998
\end{flushright}
\vspace{1.cm}

\begin{center}
\large\bf
{\LARGE\bf Testing nonperturbative techniques 
           in the scalar sector of the standard model}\\[1.5cm]
\rm
{Adrian Ghinculov$^{a,b,}$\footnote{Work supported by the 
                                  US Department of Energy (DOE)} 
 and Thomas Binoth$^c$}\\[.5cm]

{\em $^a$Randall Laboratory of Physics, University of Michigan,}\\
      {\em Ann Arbor, Michigan 48109-1120, USA}\\[.2cm]
{\em $^b$CERN, 1211 Geneva 23, Switzerland}\\[.2cm]
{\em $^c$Laboratoire d'Annecy-Le-Vieux de Physique 
         Th\'eorique\footnote{URA 1436 associ\'ee \`a l'Universit\'e de Savoie.} LAPP,}\\
      {\em Chemin de Bellevue, B.P. 110, F-74941, 
           Annecy-le-Vieux, France}\\[2.cm]

\end{center}
\normalsize

\begin{abstract}
We discuss the current picture of the standard model's scalar sector
at strong coupling. We compare the pattern observed in 
the scalar sector in perturbation theory up to two-loop
with the nonperturbative solution 
obtained by a next-to-leading order $1/N$ expansion.
In particular, we analyze two resonant Higgs scattering processes,
$f\bar f \rightarrow H \rightarrow f^{\prime} \bar{f^{\prime}}$
and $f\bar f \rightarrow H \rightarrow ZZ, WW$.
We describe the ingredients of the nonperturbative calculation,
such as the tachyonic regularization, the higher order $1/N$
intermediate renormalization, and the numerical methods for
evaluating the graphs.

We discuss briefly the perspectives and usefulness 
of extending these nonperturbative methods to other theories.
\end{abstract}

\vspace{3cm}

\end{titlepage}


\title{Testing nonperturbative techniques 
           in the scalar sector of the standard model}

\author{Adrian Ghinculov$^{a,b,}$\footnote{Work supported by the 
                                         US Department of Energy (DOE)} 
         and Thomas Binoth$^c$}

\date{{\em $^a$Randall Laboratory of Physics, University of Michigan,}\\
      {\em Ann Arbor, Michigan 48109-1120, USA}\\[.2cm]
      {\em $^b$CERN, 1211 Geneva, Switzerland}\\[.2cm]
      {\em $^c$Laboratoire d'Annecy-Le-Vieux de Physique 
         Th\'eorique\thanks{URA 1436 associ\'ee \`a l'Universit\'e de Savoie.} LAPP,}\\
      {\em Chemin de Bellevue, B.P. 110, F-74941, 
           Annecy-le-Vieux, France}}

\maketitle

\begin{abstract}
We discuss the current picture of the standard model's scalar sector
at strong coupling. We compare the pattern observed in 
the scalar sector in perturbation theory up to two-loop
with the nonperturbative solution 
obtained by a next-to-leading order $1/N$ expansion.
In particular, we analyze two resonant Higgs scattering processes,
$f\bar f \rightarrow H \rightarrow f^{\prime} \bar{f^{\prime}}$
and $f\bar f \rightarrow H \rightarrow ZZ, WW$.
We describe the ingredients of the nonperturbative calculation,
such as the tachyonic regularization, the higher order $1/N$
intermediate renormalization, and the numerical methods for
evaluating the graphs.

We discuss briefly the perspectives and usefulness 
of extending these nonperturbative methods to other theories.
\end{abstract}


\section{Introduction}

The possibility that the electroweak symmetry is broken by a
strongly interacting scalar sector received considerable 
attention in the literature. Interesting scenarios were
proposed, such as  the possibility of a Higgs boson coupled strongly
to the vector bosons and to itself \cite{veltman}, 
and the formation of a spectrum
of bound states at a higher scale which would restore unitarity in scattering
processes \cite{hveltman}. 
Also phenomenological models were proposed for studying
quantitatively the implications of strong interactions in the electroweak
symmetry breaking sector, such as the BESS model \cite{bess}.

However, beyond the phenomenological models of strong interactions,
an approach based on first principles was missing because of leak
of a nonperturbative solution, and of technical difficulties in 
extending perturbation theory in higher-loop orders.
Realistic calculations on a lattice of physical processes involving 
the Higgs sector are still confronted with technical limitations 
set, among other issues, by the
size of the lattice. $1/N$ expansions in the Higgs sector were
only performed at leading order, which is a rather poor approximation.
Perturbation theory in the Higgs sector beyond one-loop becomes very
difficult because it involves Feynman diagrams with massive internal
lines and finite external momenta, for which already at two-loop 
there are no general analytical solutions available.

Recently, considerable progress has been made in understanding from
first principles the
nature of the standard Higgs sector when its coupling becomes strong.
This is due mainly to technical advances in  massive higher-loop techniques
and in higher-order nonperturbative $1/N$ expansions. In this 
paper we would like to discuss the perturbative and nonperturbative
aspects of this behaviour at strong coupling.

The major question which will be addressed by future experiments at the
LHC is how is the electroweak symmetry broken in nature. While
it may or may not turn out to be actually broken by strong, 
nonperturbative interactions, the Higgs sector remains a fairly simple
but not trivial model, where new perturbative and nonperturbative solutions
can be tested, in view of applying them to other, 
possibly more complicated theories.


\section{Higher-order perturbation theory}

The existing calculations of leading $m_H$ radiative corrections 
in the standard Higgs sector at two-loop level
are based on using the equivalence theorem in Landau gauge. This way
radiative corrections involve only diagrams with scalars on the internal lines,
so that the problem at hand becomes much simpler. This procedure was
proposed for the first time in ref. \cite{marciano}, 
where the one-loop correction
to Higgs decay into vector bosons was calculated in this way.

So far at two-loop level the scalar self-energies and the main decay modes
of a heavy Higgs boson ($t\bar{t}$, WW, ZZ) are known, and also the
high energy limit of vector boson scattering 
\cite{calc2loop:all}. 
The vector boson scattering 
is known completely only at one-loop level \cite{wwscat:all}. 
This is due to the 
complexity of the diagrams involved in a complete two-loop treatment. 
By using
the existing two- and three-point functions, some other scattering
processes of phenomenological interest can be derived 
\cite{mucollider,gluefusion}.
For a discussion of the existing results concerning
effects of enhanced electroweak strength in the standard model 
at two-loop order, see for instance ref. \cite{2loop:review}.

The main decay modes of heavy Higgs bosons are into pairs of vector bosons
and into top quark pairs. At leading order, these decay width are given by the
following expressions:

\begin{eqnarray}
\Gamma^{(tree)}_{H \rightarrow t \bar{t}} & = & 
 \frac{3 g^2}{32 \pi} \frac{m_H \, m_t^2}{m_W^2} 
 \left( 1 - 4 \frac{m_t^2}{m_H^2} \right)^{3/2}  \; \; ,
    \nonumber \\
\Gamma^{(tree)}_{H \rightarrow W^+ W^-} & = & 
 \frac{g^2}{64 \pi} \frac{m_H^3}{m_W^2} 
 \left( 1 - 4 \frac{m_W^2}{m_H^2} \right)^{1/2}  
 \left( 1 - 4 \frac{m_W^2}{m_H^2} + 12 \frac{m_W^4}{m_H^4} \right) \; \; ,
    \nonumber \\
\Gamma^{(tree)}_{H \rightarrow Z^0 Z^0} & = & 
 \frac{g^2}{128 \pi} \frac{m_H^3}{m_W^2} 
 \left( 1 - 4 \frac{m_Z^2}{m_H^2} \right)^{1/2}  
 \left( 1 - 4 \frac{m_Z^2}{m_H^2} + 12 \frac{m_Z^4}{m_H^4} \right)
      \; \; .
\end{eqnarray}

The radiative corrections of enhanced electroweak strength up to two-loop
order are given by the following multiplicative factors \cite{calc2loop:all}:

\begin{eqnarray}
\lefteqn{ \Gamma_{H \rightarrow t \bar{t}} \, =  \,
                      \Gamma^{(tree)}_{H \rightarrow t \bar{t}} \, \times 
     \left[
 1 + \lambda 
     \left(  \frac{13}{8} - \frac{\pi \, \sqrt{3}}{4} \right)  
   - \lambda^2
     \left( \, .51023 \pm 2.5 \cdot 10^{-4} \, \right)
 \right]
 }
 \nonumber \\
 & = & \Gamma^{(tree)}_{H \rightarrow t \bar{t}} \,
  \left[
 1 + .264650 \, \lambda 
   - \left( \, .51023 \pm 2.5 \cdot 10^{-4} \, \right) \lambda^2
 \right]
 \nonumber \\
\lefteqn{ \Gamma_{H \rightarrow W^+ W^- \, , \, Z^0 Z^0} \, =  \,
          \Gamma^{(tree)}_{H \rightarrow W^+ W^- \, , \, Z^0 Z^0} \, \times }
 \nonumber \\
 & &  \left[
 1 + \lambda 
     \left( \frac{19}{8} + \frac{5 \, \pi^2}{24} 
          - \frac{3 \, \sqrt{3} \, \pi}{4} 
    \right)  
   + \lambda^2
     \left( \, .97103 \pm 8.2 \cdot 10^{-4} \, \right)
 \right]
    \nonumber \\
 & = & \Gamma^{(tree)}_{H \rightarrow W^+ W^- \, , \, Z^0 Z^0} \,
 \left[
 1 + .350119 \, \lambda 
   + \left( \, .97103 \pm 8.2 \cdot 10^{-4} \, \right) \lambda^2
 \right]
      \; \; .
\end{eqnarray}
Here $\lambda = (\frac{g}{4 \pi} \frac{m_{H}}{m_{W}})^2$
is the quartic coupling of the scalar sector.

\begin{figure}
\hspace{1.cm}
    \epsfxsize = 14.5cm
    \epsffile{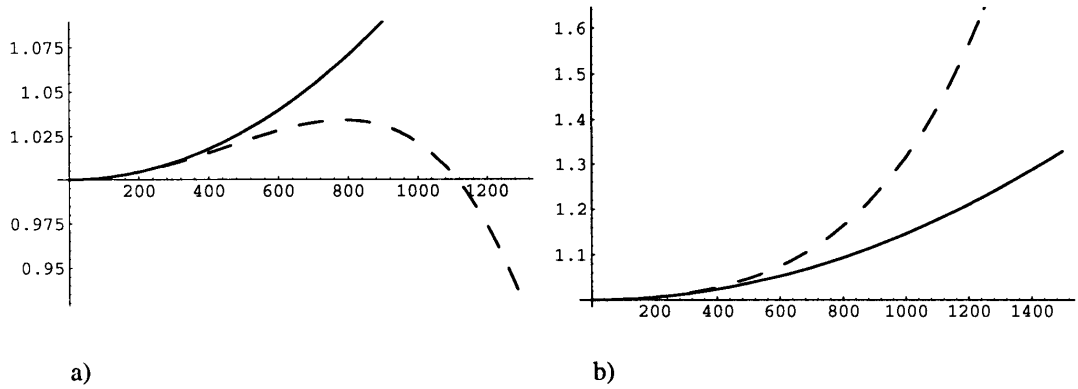}
\caption{{\em The magnitude of the leading $m_H$ radiative corrections to
         the $H \rightarrow t\bar{t}$ (a) and 
         the $H \rightarrow ZZ, WW$ (b) decays. The plots show the ratios
         of the decay widths at one-loop (solid line) and two-loop
         (dashed line) to the tree level decay widths as a function of
         the on-shell Higgs mass.}}
\end{figure}

In the above expressions, the strength of the coupling of the scalar sector
is parameterized by the on-shell Higgs mass $m_H$, which is defined by the
on-shell renormalization condition 
$Re \left[P_{HH}^{-1}(s=m_H^2)\right]=0$. We plot
the correction factors given by eqns. 2 in figure 1.
One can see that for both correction factors the two-loop correction 
becomes as large as the one-loop correction for $m_H$ about 1 TeV (1.1 TeV 
for $H \rightarrow f\bar{f}$, and 930 GeV for $H \rightarrow WW, ZZ$).
Even if the higher-loop corrections are not yet known, this pattern
suggests that the radiative corrections blow up 
strongly around $m_H \sim$ 1 TeV.
Note that the scheme ambiguities associated with the two-loop result
become substantial already for considerable smaller values of the
Higgs mass \cite{riess:ms}.
Of course, similar conclusions can be obtained from other scattering
processes apart from the decay modes.
This behaviour is in agreement with well-established results 
regarding perturbative
unitarity violation in vector boson scattering 
at tree level \cite{lee-quigg}.

The point where the perturbative expansion blows up depends on the
expansion parameter, and therefore on the renormalization scheme. 
So far the two-loop results mentioned above
were translated into the $\overline{MS}$ \cite{riess:ms} 
and the pole \cite{mucollider,higgspole} 
renormalization schemes for processes at the Higgs mass
energy scale, in the hope that perturbation theory
may show better convergence properties in certain schemes.
An overall conclusion of these
studies is that the $\overline{MS}$ scheme diverges somewhat
sooner and has larger scheme uncertainty than the on-shell
scheme, while the pole scheme appears to have slightly better
convergence properties than the on-shell scheme.
For a more refined discussion of the scheme and gauge dependence
see ref. \cite{kniehlsirlin}

While the existing higher-order perturbative results in the scalar sector 
are consistent with a strong 
blow-up of radiative corrections at about $m_H \sim$
1 TeV, this fact appears somehow puzzling if one considers that the
quartic coupling of the scalar sector, 
$\lambda = (\frac{g}{4 \pi} \frac{m_{H}}{m_{W}})^2$, 
is only of order .4 for $m_H \sim$ 1 TeV.

From the perspective of the nonperturbative solution to
be discussed in the following section,
the reason for this behaviour is the Higgs mass saturation effect. 
Due to the dynamics of the scalar sector, the on-shell Higgs mass 
is not a good parameterization of the quartic coupling for values
larger than $m_H \sim$ 1 TeV. However, as will be shown in the
following section, at the fundamental level,
the scalar sector is perfectly well defined at
higher values of the quartic coupling. Well-behaved, unitary
solutions can be obtained for various processes by using 
nonperturbative methods. 

At a more speculative level,
the fact that the pole scheme seems to show somewhat better 
convergence properties than the on-shell scheme may have
its origin in another nonperturbative property of the 
Higgs self-energy. This type of nonperturbative effect 
was discussed in ref. \cite{beenakker} in the context of a theory of
vector bosons coupled to a large number of light fermions.
The idea is that due to the structure of the self-energy
function the on-shell renormalization condition may have
no solution for a range of the theory's coupling, while 
the pole renormalization condition may have a solution.
This results in the on-shell mass being suitable for 
parameterizing the theory only for a more limited range of 
the physical coupling than the pole mass. Should such
a nonperturbative effect be present in the Higgs sector,
this could explain the pattern observed in perturbation theory.
In ref. \cite{higgspole} this was investigated by using
the nonperturbative $1/N$ expansion, and it was shown
that this effect is not present in the Higgs sector of
the standard model at LO. 
It is still an open question if it appears at NLO.


\section{Nonperturbative $1/N$ solution}

The $1/N$ expansion aims at a nonperturbative solution in order to avoid 
the problems of perturbation theory at large coupling. Perturbation 
theory ceases to be a satisfactory solution when radiative correction
blow up already at lower loop orders and the renormalization scheme 
ambiguity is so large that the result becomes unreliable.

The $1/N$ approach is free of these problems because all radiative
corrections of all loop orders are explicitly summed up. The idea
is to treat the Higgs sector as an $O(N)$-symmetric sigma
model, where $N=4$ for the standard model, and to expand in $1/N$
instead of the quartic coupling. The solution 
is then valid independently of the strength of the coupling --- the
quality of the approximation depends on the value of $N$. Also
it is completely free of renormalization scheme ambiguities. One
can work in any intermediate renormalization scheme and still obtain
the same result.

Another interesting nonperturbative feature of the $1/N$ expansion
of the sigma model is the finiteness of 
wave function renormalization constants.
This is a property of the exact nonperturbative solution, 
and was checked at next-to-leading order in $1/N$. However,
the renormalization of the coupling constants is ultraviolet
divergent.

As an extra bonus, the $1/N$ solution provides naturally a 
consistent treatment of resonant scattering amplitudes. This was known 
as a long-standing issue in perturbation theory. The essence
of the problem is that around a resonance one has to perform a
Dyson summation which in perturbation theory at any finite
order introduces incomplete higher order contributions which are unphysical.
In gauge theories this leads to gauge dependent results. Meanwhile
a solution to this problem was found \cite{stuart:1,stuart:2},
which is based on a Laurent expansion around the physical pole. 
Within perturbation theory, 
this solves the problem in a fundamental way, and applies 
consistently to all orders in perturbation theory. 
Another approach which was proposed is to use gauge invariant
pinch technique self-energies in the Dyson summation.
For phenomenological purposes only, other
approaches to treat 
resonant amplitudes were proposed in the literature, 
which amount to special resummations of low oder results.
Some of them may be easier to use in certain phenomenological applications.
At the same time they are less fundamental theoretically --- some of them 
still contain unphysical higher order terms, or apply only
to tree level or one-loop calculations.

The $1/N$ solution is automatically free of these problems
of perturbation theory because it is an all-order solution
in the loop expansion.

However, the $1/N$ solution still has some residual ambiguity
which can be related to the triviality problem and to possible
nondecoupling effects from a hidden heavy sector. Technically,
this appears in the tachyonic regularization. Perturbation
theory is used at an intermediary stage in the usual $1/N$ treatment,
but perturbation theory does not determine the solution uniquely. 
This is the physical origin of the ambiguity entailed 
in the tachyonic regularization.

\subsection{$1/N$ combinatorial rearrangement and diagrammatics}

We start with the usual Lagrangean of a $O(N)$-symmetric
sigma model:

\begin{equation}
  {\cal L}_1 =   \frac{1}{2}           
               \partial_{\nu}\Phi_0 \partial^{\nu}\Phi_0 
             - \frac{\mu_0^2}{2}      \Phi_0^2 
	     - \frac{\lambda_0}{4! N} \Phi_0^4 
 ~~ , ~~ \Phi_0 \equiv \left( \phi_0^1, \phi_0^2, \dots , \phi_0^N \right)
\end{equation}

From this Lagrangean one can in principle derive directly
a perturbative expansion for Green functions, and classify
the Feynman diagrams according to their order in $1/N$. However,
beyond leading order the combinatorics becomes very complicated.
In order to perform explicit calculations beyond leading order
it is useful to perform a rearrangement of perturbation theory.
A useful trick for doing this was proposed in ref. \cite{coleman}.
It consists of adding a nondynamical term to the Lagrangean:

\begin{eqnarray}
  {\cal L}_2 & = & {\cal L}_1 + \frac {3 N}{2 \lambda_0} 
                 (\chi_0 - \frac{\lambda_0}{6 N} \Phi_0^2 - \mu_0^2)^2 
	\nonumber \\	
           & = &				     
    \frac{1}{2} \partial_{\nu}\Phi_0 \partial^{\nu}\Phi_0 
  - \frac{1}{2} \chi_0 \Phi_0^2 
  + \frac{3 N}{2 \lambda_0} \chi_0^2
  - \frac{3 \mu_0^2 N}{\lambda_0} \chi_0 + const. 
\end{eqnarray}

This involves the introduction of an unphysical auxiliary field
$\chi$. As one can see, the equation of motion for $\chi$ is
just an equation of constraint, and therefore the physical spectrum
and the dynamics of the model remain unchanged. The effect of
this trick is that the Feynman rules are changed. Namely, the
quartic couplings are eliminated. The only vertices left are
trilinear, and involve one $\chi$ field and two physical scalars.
This simplifies enormously the combinatorics of Feynman graphs
in higher orders.

We note that some other rearrangement schemes with different
properties were discussed for the $O(N)$-symmetric sigma 
model in ref. \cite{root}. To our best knowledge, these 
schemes were not applied so far in actual calculations.

In the following we describe the counterterm structure
which is used for performing renormalization at NLO in
the $1/N$ expansion. A somewhat different approach is
presented in ref. \cite{thomas:rev}, but the final results are
the same.

Renormalization is performed in principle order by order
in perturbation theory. However, for performing actual
calculations of higher order in $1/N$, 
it is of advantage to group all counterterms
of various loop orders and which are of the same $1/N$ order 
into the same $1/N$ counterterm. 
We define the $1/N$ counterterms as follows:

\begin{eqnarray}
  \frac{3}{\lambda_0}       &=&   \frac{3}{\lambda} + \Delta \lambda   
  ~~~~~~~~~~~ \equiv ~~ \frac{3}{\lambda} + \delta \lambda^{(0)} 
                             + \frac{1}{N} \delta \lambda
                             + {\cal O}\left(\frac{1}{N^2}\right)
\nonumber \\
  \frac{3 \mu_0}{\lambda_0} &=& - \frac{v^2}{2} (1 + \Delta t)
  ~~~ \equiv ~~ - \frac{v^2}{2} \left[ 1 + \frac{1}{N}\delta t 
                               + {\cal O}\left(\frac{1}{N^2}\right) \right]
\nonumber \\
  \phi^i_0 &=&  \pi_i  Z_{\pi}          
  ~~~~~~~~~~~~~~~\, \equiv ~~ \pi_i \left[ 1 + \frac{1}{N} \delta Z_{\pi}
                          + {\cal O}\left(\frac{1}{N^2}\right)  \right] 
 ~ , ~~  i=1, \dots ,N-1 
\nonumber \\
  \phi^N_0 &=&  \sigma Z_{\sigma} + \sqrt{N} v                         
  ~~~~~~ \equiv ~~ \sigma \left[ 1 + \frac{1}{N} \delta Z_{\sigma}
                          + {\cal O}\left(\frac{1}{N^2}\right)  \right] 
         + \sqrt{N} v
\nonumber \\
  \chi_0   &=&  \chi   Z_{\chi} + \hat \chi +\Delta t_{\chi}
  ~~ \equiv ~~ \chi \left( 1 + \frac{1}{N} \delta Z_{\chi}  \right) 
          + \frac{v^2}{N} \delta t_{\chi} + {\cal O}\left(\frac{1}{N^2}\right)
\end{eqnarray}
Here we already used the fact that the tadpole and wave function 
renormalization counterterms do not receive contributions 
at leading order in $1/N$. We also note that although two tadpole counterterms
are present, $\delta t$ and $\delta t_{\chi}$, they are related
through the gap equation \cite{bardeen}, for instance by requesting that the
leading order ground state condition $\hat{\chi} = 0$ be preserved
in higher orders, where $\hat{\chi}$ is the vacuum expectation value
of the $\chi$ field in the spontaneously broken phase.

At this point it is useful to note that since the two Lagrangeans
of eqns. 3 and 4 are equivalent, a linear combination will also
describe the same physics. This observation can be exploited for
performing BPHZ renormalization in a more elegant way, as will 
be explained in the following. Beyond leading order in $1/N$ 
it is advantageous to work with a linear combination of the
potential parts of Lagrangeans ${\cal L}_1$ and ${\cal L}_2$.
Keeping only the contributions relevant for next-to-leading
order calculations, we consider in fact the following Lagrangean:

\begin{eqnarray}
  {\cal L}_3 & = & \frac {1}{2} (\partial_{\mu} \vec\pi)
                                (\partial^{\mu} \vec\pi) 
                 \left[ 1 + \frac{2}{N}\delta Z_{\pi} \right]
               + \frac {1}{2} (\partial_{\mu} \sigma)(\partial^{\mu} \sigma) 
                 \left[ 1 + \frac{2}{N}\delta Z_{\sigma} \right]
	\nonumber \\	
           &  &				     
     + \chi^2 \frac{N}{2} \left\{ \frac{3}{\lambda}
            + \delta \lambda^{(0)} 
            + \frac{1}{N} \left[ \delta \lambda 
            + 2 \left( \frac{3}{\lambda} + \delta \lambda^{(0)} \right)
                \delta Z_{\chi} \right]  \right\}
	\nonumber \\	
           &  &				     
     -  \chi \sigma \sqrt{N} v \left[ 1 + \frac{1}{N} 
                  \left( \delta Z_{\chi} + \delta Z_{\sigma} \right) \right]
	\nonumber \\	
           &  &				     
     - \frac{1}{2} \left( \vec\pi^2 + \sigma^2 \right)
                   \frac{v^2}{N} \delta t_{\chi}
	\nonumber \\	
           &  &				     
     - \frac{1}{2} \chi \vec\pi^2 \left[ 1 
       + \frac{1}{N} \left( \delta Z_{\chi} + 2 \delta Z_{\pi} \right)  \right]
	\nonumber \\	
           &  &				     
     - \frac{1}{2} \chi \sigma^2 \left[ 1 
       + \frac{1}{N} \left( \delta Z_{\chi} 
                        + 2 \delta Z_{\sigma} \right)  \right]
	\nonumber \\	
           &  &				     
     - \sigma \frac{v^3}{\sqrt{N}} \delta t_{\chi}
     + \chi v^2 \left[ \left( \frac{3}{\lambda} + \delta \lambda^{(0)} \right)
                       \delta  t_{\chi} + \frac{1}{2} \delta t
                     + \frac{1}{2N} \delta t \; \delta Z_{\chi} \right]
	\nonumber \\	
           &  &				     
     + \frac{K}{N^2} \left[ 4 N v^2 \sigma^2 
                           + \left( \vec\pi^2 \right)^2 + \sigma^4
 + 2 \vec\pi^2 \sigma^2 
 + 4\sqrt{N}v\sigma \left( \vec\pi^2 + \sigma^2 \right) \right]
\end{eqnarray}
Here $K$ is in principle a completely arbitrary constant.  
We have the freedom to choose it so that actual calculations
are more convenient. We will consider
$K$ to be of order 1 in the $1/N$ expansion. Thus the potential 
part of ${\cal L}_2$ is regarded as an ${\cal O}(1/N)$ counterterm.
We will choose the actual value of $K$ so that the renormalization 
procedure is more transparent at NLO in the $1/N$ expansion.

The Feynman rules can be read out directly from the above expression.
Within this diagrammatical rearrangement of the sigma model, 
counting powers of $1/N$ in multiloop diagrams is straightforward:   
closed Goldstone loops contribute a factor $N$, while $\chi \chi$
propagators give a $1/N$ factor, and mixed $\chi \sigma$ propagators
contribute $1/\sqrt{N}$. At the same time, the absence of quartic
couplings at tree level reduces considerably the number of
possible topologies. 

For these reasons, for a given process
and a given order in $1/N$, it is easy to write down the Feynman graphs
of all loop orders. As it will become clear from the discussion of
two- and three-point functions, there is always a finite number 
of multiloop topologies, where one can only insert chains of one-loop
bubbles in the $\chi \chi$ and $\chi \sigma$ propagators without
increasing the $1/N$ order of the graph.

We emphasize that this combinatorial rearrangement of the sigma model
is quite crucial. It is possible to calculate explicitly nonperturbative
processes in the Higgs sector precisely because the combinatorial
rearrangement enables one to write down explicitly and in a
manageable way the diagrams of all loop orders, without truncating
the perturbative expansion.

\subsection{Tachyonic regularization}

It is straightforward to derive the two-point functions of the theory
at leading order in $1/N$. This was done for instance in ref. 
\cite{coleman}. The only diagram involved is the one-loop 
bubble diagram shown in figure 2. One finds the following leading
order propagators:

\begin{figure}
\begin{tabular}{ccccc}
 \hspace{2cm}
 $i (N-1) \hat{\alpha}^{(0)}(s)$ & $=$ & 
    \raisebox{-.74cm}{\epsfxsize = 3.3cm \epsffile{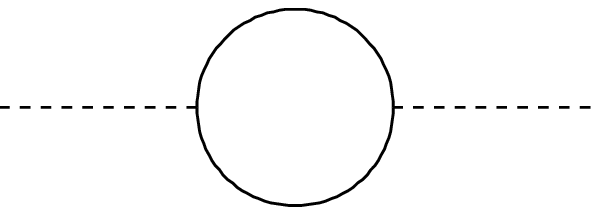}} & + & counterterm 
\end{tabular}
\caption{{\em The leading order bubble diagram.}}
\end{figure}

\begin{eqnarray}
  D_{\sigma \sigma}(s) &=&                        \frac{i         }{s - m^2(s)} \nonumber \\
  D_{\chi \chi}(s)     &=& \frac{1}{      N  v^2} \frac{i s m^2(s)}{s - m^2(s)} \nonumber \\
  D_{\chi \sigma}(s)   &=& \frac{1}{\sqrt{N} v  } \frac{i   m^2(s)}{s - m^2(s)} \nonumber \\
  D_{\pi_i \pi_j}(s)   &=& \frac{i}{s}  \delta_{ij}  
\end{eqnarray}
where

\begin{equation}
  m^2(s) = \frac {v^2}{\frac{3}{\lambda} + \hat{\alpha}^{(0)}(s)} 
    ~~~~~  \equiv  ~~~~~ 
            \frac {v^2}{\frac{3}{\lambda} - \frac{1}{32 \pi^2} 
                        \log{\left(-\frac{s+i \eta}{\mu^2}\right)}} 
\end{equation}     
Here $\hat{\alpha}^{(0)}(s) \equiv \alpha^{(0)}(s) + \delta \lambda^{(0)}$ 
is the ultraviolet finite part of the self-energy diagram of fig. 1,
and $\mu$ is the subtraction scale.

In the expressions above, apart from the expected Higgs pole, 
one notices the presence of a tachyonic pole. It appears at an 
energy $s=-\Lambda_t^2$ which is given by the following 
transcendental equation:

\begin{equation}
    \frac{v^2}{\Lambda_t^2} 
  - \frac{1}{32 \pi^2} \log{\left( \frac{\Lambda_t^2}{\mu^2} \right)}
  + \frac{3}{\lambda} = 0   ~~~~~ .
\end{equation}
The tachyon scale $\Lambda_t$ differs from the
Landau scale $\Lambda_L = \mu e^{48 \pi^2/\lambda}$ by a shift
of order $v^2/\Lambda_L^2$.

The leading order tachyon is a well-known difficulty of the $1/N$
nonperturbative treatment of the sigma model. From the technical point
of view, it induces causality violating effects in the theory.
As long as one is concerned only with the leading order, one can try
to make sense of the result by limiting its validity to an energy
range considerably smaller than the tachyon scale. However, there 
is no such easy way out for calculations beyond leading order
because the tachyon appears then in loops.

One way to circumvent this is by making the assumption that the tachyon 
indicates the triviality of the theory. This sets a limit for the validity
of the $1/N$ result at high energy. At some energy scale new physics
sets in. This scale is presumably 
of the order of the tachyon scale, but not necessarily equal or lower.
Then an obvious treatment is to introduce a cutoff in the loop
integrations \cite{nunes}. 
This is then interpreted as a model of nondecoupling
effects from an unknown heavy sector. However, in this approach the 
momentum cutoff has to be lower than the tachyon scale, which is necessary
for computational purposes only and is not motivated physically. Also a
loop momentum cutoff spoils the gauge invariance of the gauged
model. It also introduces quadratic dependencies on the cutoff
scale and these are known in effective theories not to be directly
related to heavy mass effects --- actually counterexamples 
were found in the literature in two-loop calculations \cite{vdbij:rho}.

We use a different treatment of the tachyonic pole \cite{oneovn}, which is
more convenient for higher order calculations in the $1/N$
expansion. We subtract the tachyon minimally at its pole,
which means using the following propagators instead of those
of eqns. 7:

\begin{eqnarray}
  D_{\sigma \sigma}(s) &=& i \left[   \frac{1     }{s - m^2(s)}
                                 - \frac{\kappa}{s + \Lambda_t^2} \right] \nonumber \\
  D_{\chi \chi}(s)     &=& \frac{i s}{N  v^2} 
                          \left[   \frac{m^2(s)     }{s - m^2(s)}
                                 + \frac{\kappa \Lambda_t^2}{s + \Lambda_t^2} \right] \nonumber \\
  D_{\chi \sigma}(s)   &=& \frac{i}{\sqrt{N} v}  
                          \left[   \frac{m^2(s)     }{s - m^2(s)}
                                 + \frac{\kappa \Lambda_t^2}{s + \Lambda_t^2} \right]
 ~~~~~,
\end{eqnarray}
were 

\begin{equation}
  \kappa =  \frac{1}{ 1 + \frac{\Lambda_t^2}{32 \pi^2 v^2} }
\end{equation}
is the residuum of the tachyonic pole. 

The justification of the tachyonic regularization introduced above
is the following. Green functions --- such as the two-point functions
above ---  are calculated in the $1/N$ expansion starting with the
perturbative expansion in the coupling constant $\lambda$ of
the coefficients of the $1/N$ expansion. Then, all Feynman graphs
of all loop orders which contribute to a given order in $1/N$ 
are calculated explicitly and summed up. Since the $1/N$ coefficient
is only known as a power series in $\lambda$ to start with, it will
be determined by its perturbative expansion up to a function of
$\lambda$ which vanishes identically in perturbation theory,
of the type $e^{1/\lambda}$. Since the residuum of the tachyon
is such a function whose perturbative expansion vanishes, 
its presence cannot be taken seriously as a prediction of the 
theory and as an indication that the theory is ill-defined.
While the $\lambda \phi^4$ theory is widely believed to be
trivial, the tachyon in the $1/N$ expansion is certainly not
a rigorous proof thereof.

When we use the minimal tachyon subtraction scheme of eqns. 10,
we effectively use the freedom to add a function which vanishes in
perturbation theory for restoring the causality of the theory.
In this sense, our treatment is independent of whether one regards
the theory as being an effective one or not. If one takes the view
that the theory is trivial and wants to include nondecoupling effects
from a heavy sector, such effects can be superimposed over the whole
calculation. Compared to a naive cutoff, this tachyonic regularization
does not require the heavy sector to be strictly under the tachyon scale.

\subsection{Nonperturbative two- and three-point functions at NLO}

Beyond leading order in $1/N$, actual calculations have to
be performed numerically because in general the multiloop diagrams involved
are not manageable analytically. This brings about some technical
complications related to the treatment of ultraviolet divergencies
in conjunction with numerical integration.

A useful observation is that the final result in the $1/N$ expansion
is free of any renormalization scheme ambiguity. This is because
it is exact at all orders in the coupling constant. This leaves
us the freedom of working in any intermediate renormalization scheme
at our convenience, since the final result is independent of that.
This can be best exploited for simplifying to some extent the
numerical work.

\begin{figure}
\begin{tabular}{cccccccc}
      $i \alpha(s)$
    & $=$
    & 
    &       \raisebox{-.74cm}{$A_1$ \hspace{-1cm} \epsfxsize = 3.3cm \epsffile{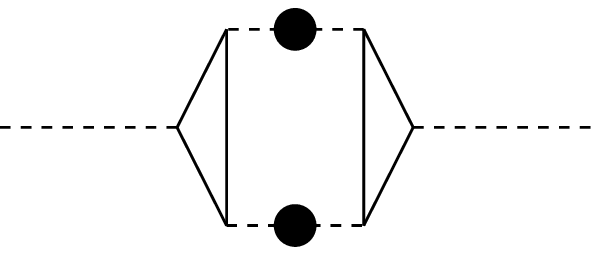} }  
    & $+$ & \raisebox{-.74cm}{$A_2$ \hspace{-1cm} \epsfxsize = 3.3cm \epsffile{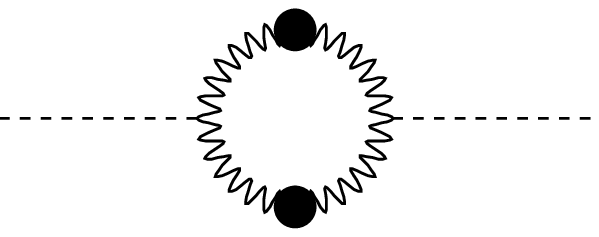} }  
    & $+$ & \raisebox{-.74cm}{$A_3$ \hspace{-1cm} \epsfxsize = 3.3cm \epsffile{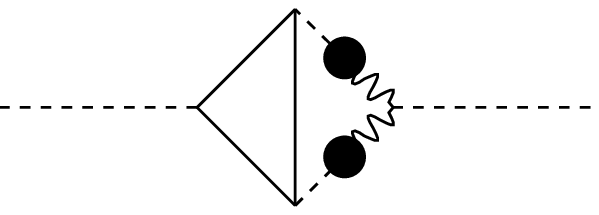} } 
  \\
    &  
    & $+$ & \raisebox{-.74cm}{$A_4$ \hspace{-1cm} \epsfxsize = 3.3cm \epsffile{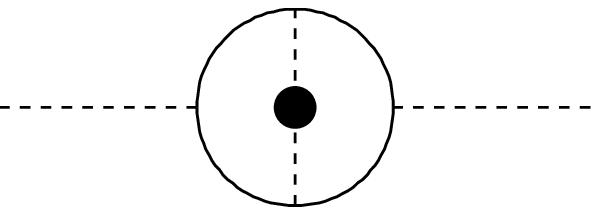} }  
    & $+$ & \raisebox{-.74cm}{$A_5$ \hspace{-1cm} \epsfxsize = 3.3cm \epsffile{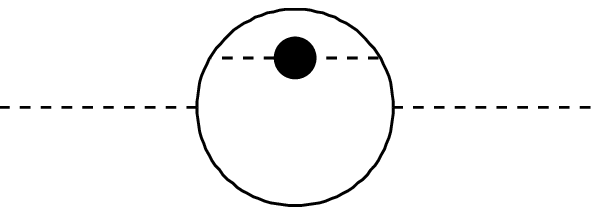} }
    & 
    & 
  \\
      $i \frac{v^2}{N} \beta(s)$
    & $=$
    & 
    &       \raisebox{-.74cm}{$B_1$ \hspace{-1cm} \epsfxsize = 3.3cm \epsffile{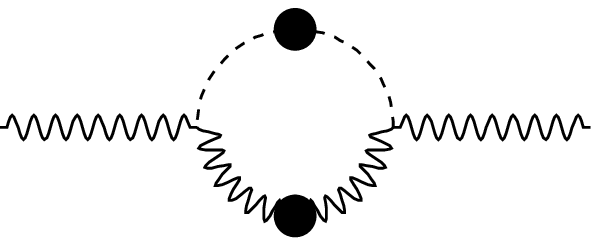} }  
    & $+$ & \raisebox{-.74cm}{$B_2$ \hspace{-1cm} \epsfxsize = 3.3cm \epsffile{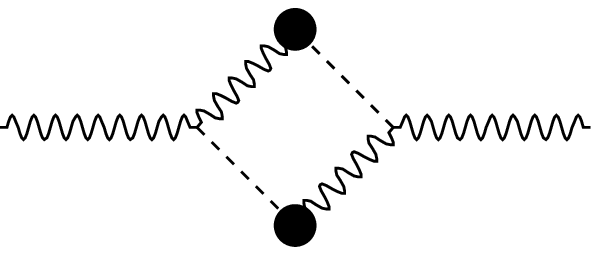} } 
    &   
    & 
  \\
      $i \frac{v}{\sqrt{N}} \gamma(s)$
    & $=$
    & 
    &       \raisebox{-.74cm}{$C_1$ \hspace{-1cm} \epsfxsize = 3.3cm \epsffile{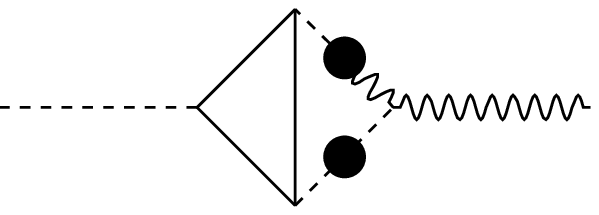} }  
    & $+$ & \raisebox{-.74cm}{$C_2$ \hspace{-1cm} \epsfxsize = 3.3cm \epsffile{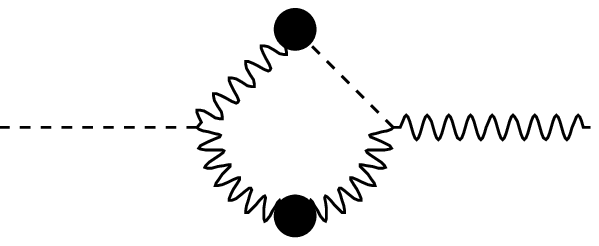} } 
    & 
    & 
  \\
      $i \frac{v^2}{N} \delta(s)$
    & $=$
    & 
    & \raisebox{-.74cm}{$D$ \hspace{-1cm} \epsfxsize = 3.3cm \epsffile{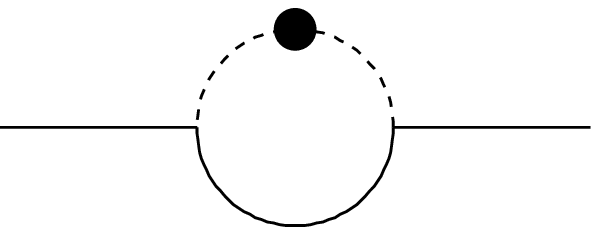} }
    &   
    & 
    & 
    & 
  \\
    T
    & $=$
    & 
    & \raisebox{-1.cm}{ \hspace{-1cm} \epsfxsize = 3.3cm \epsffile{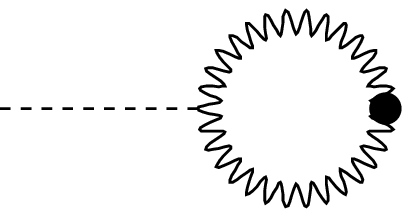} }
    &  
    & 
    & 
    & \raisebox{-.74cm}{\epsfxsize = 3.3cm \epsffile{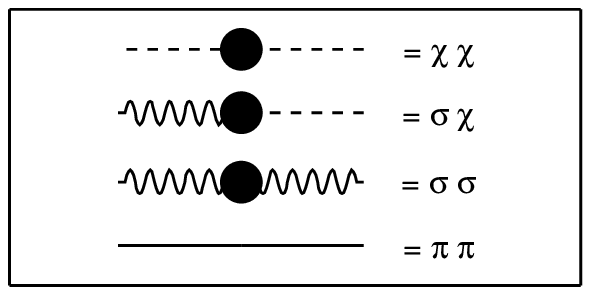}}
  \\
    - $i \frac{v}{N\sqrt{N}} \eta(s)$
    & $=$
    & 
    & \raisebox{-.74cm}{$E_1$ \hspace{-1cm} \epsfxsize = 3.3cm \epsffile{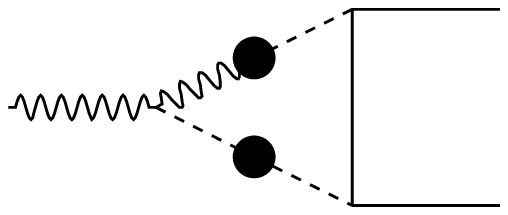} }
    &  
    & 
    & 
    & 
  \\
     - $i \frac{1}{N} \phi(s)$ 
    & $=$
    & 
    &       \raisebox{-.74cm}{$F_1$ \hspace{-1cm} \epsfxsize = 3.3cm \epsffile{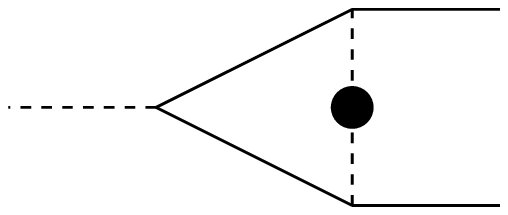} }  
    & $+$ & \raisebox{-.74cm}{$F_2$ \hspace{-1cm} \epsfxsize = 3.3cm \epsffile{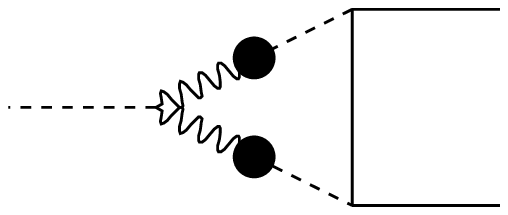} }  
    & $+$ & \raisebox{-.74cm}{$F_3$ \hspace{-1cm} \epsfxsize = 3.3cm \epsffile{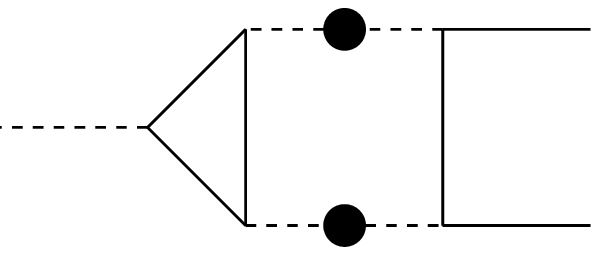} }
\end{tabular}
\caption{{\em Infinite sums of multiloop Feynman diagrams which contribute
              in next-to-leading order in $1/N$
              to the two- and three-point functions of the $O(N)$ sigma model.
	      The blob on propagators denotes the summed-up leading order
	      propagators. Note that the $\pi \pi$ propagator at leading
	      order in $1/N$ is a free propagator.
	      One of the graphs above is shown in expanded form in fig. 4.}}
\end{figure}

The graphs needed for the calculation of all two- and three-point
functions of the theory at next-to-leading order in $1/N$ 
are shown in fig. 3. In addition to the two- and three-point
graphs, there is also one tadpole graph
which is needed for the determination of the tadpole
counterterm $\delta t_{\chi}$. Each graph is in fact a sum of multiloop
Feynman graphs which are all of the same order in $1/N$, and of
various orders in the coupling constant $\lambda$. This is shown explicitly
for one particular self-energy graph in fig. 4.

\begin{figure}
\begin{tabular}{cccc}
      $i A_1(s) =$
    &        \raisebox{-.99cm}{\epsfxsize = 4.4cm \epsffile{a6.eps}}  
    & \parbox{2cm}{ \[ \equiv ~~ \sum_{k = 0}^{\infty}
                                 \sum_{l = 0}^{\infty} \] }
    & \raisebox{-1.81cm}{\epsfxsize = 8cm \epsffile{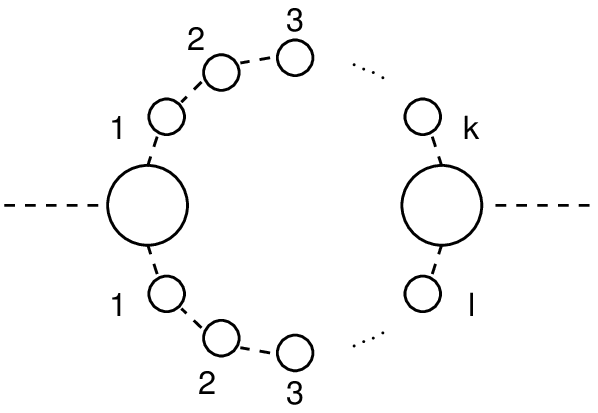}}  
\end{tabular}
\caption{{\em Multiloop diagrams with three-loop topology which
              contribute to the $\chi \chi$ propagator 
	      in next-to-leading order.}}
\end{figure}

The graphs involved, $A_i$, $B_i$, $C_i$, $D_i$, $E_i$, $F_i$ and $T$,
are in general ultraviolet divergent. Since our strategy is to calculate
them numerically, we first subtract the divergencies and subdivergencies
of these graphs. An inspection of the Feynman diagrams which compose
each graph of figure 3 reveals that the ultraviolet divergencies are 
polynomial, just as if they were usual Feynman diagrams, and 
in spite of the infinite number of loops involved. We define 
in figures 5, 6 and 7 a set of ultraviolet subtracted graphs,
$\hat{A}_i$, $\hat{B}_i$, $\hat{C}_i$, $\hat{D}_i$, 
$\hat{E}_i$, and $\hat{F}_i$.
They are finite, and thus can be calculated by 
direct numerical integration.

\begin{figure}
\begin{tabular}{rcccccl}
      $i \hat{A}_4(s) ~=$
    & 
    &       \raisebox{-.74cm}{ \epsfxsize = 3.3cm \epsffile{a6.eps} }  
    & $-$ & \raisebox{-.74cm}{ \epsfxsize = 3.3cm \epsffile{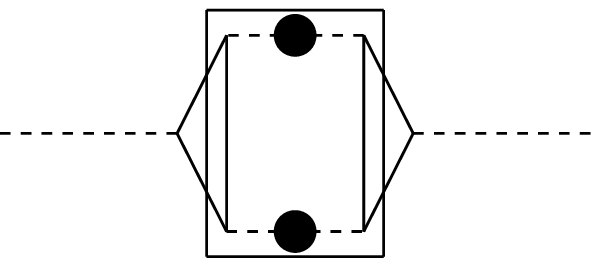} }  
    & 
    &  
  \\
    & $- ~2 \times$
    &       \raisebox{-.74cm}{ \epsfxsize = 3.3cm \epsffile{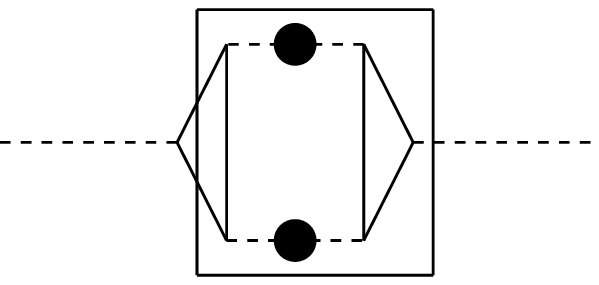} }  
    & $+ ~2 \times$ & \raisebox{-.74cm}{ \epsfxsize = 3.3cm \epsffile{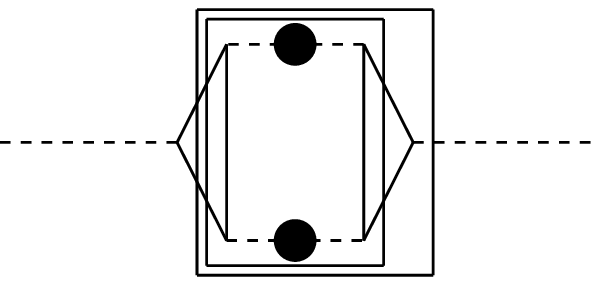} }  
    & $-$
    & overall subtraction
  \\
      $i \hat{A}_2(s) ~=$
    & 
    &       \raisebox{-.74cm}{  \epsfxsize = 3.3cm \epsffile{a2.eps} }  
    & $-$ & \raisebox{-.74cm}{  \epsfxsize = 3.3cm \epsffile{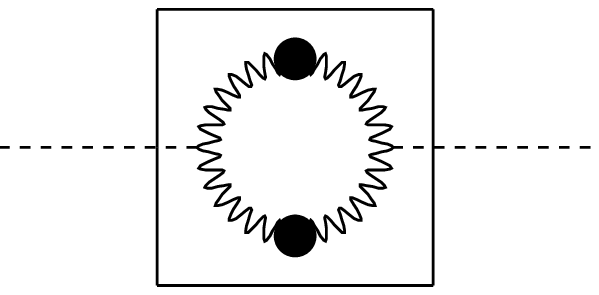} }  
    & 
    & 
  \\
      $i \hat{A}_3(s) ~=$
    & 
    &       \raisebox{-.74cm}{ \epsfxsize = 3.3cm \epsffile{a3.eps} }  
    & $-$ & \raisebox{-.74cm}{ \epsfxsize = 3.3cm \epsffile{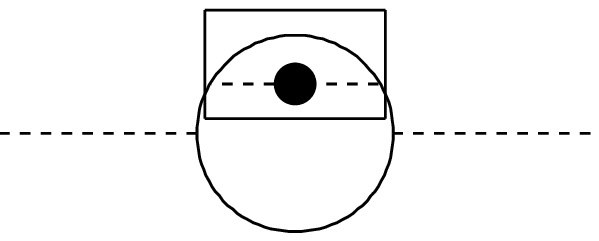} }  
    & 
    & 
  \\
    & $-$ & \hspace{-1.7cm}\raisebox{-.74cm}{ \epsfxsize = 3.3cm \epsffile{a3c1.eps} \hspace{-2.25cm} \raisebox{1.15cm}{$\partial$} } 
    & $-$ & overall subtraction
    & 
    & 
  \\
      $i \hat{A}_4(s) ~=$
    & 
    &       \raisebox{-.74cm}{ \epsfxsize = 3.3cm \epsffile{a4.eps} }  
    & $- ~2 \times$ & \raisebox{-.74cm}{ \epsfxsize = 3.3cm \epsffile{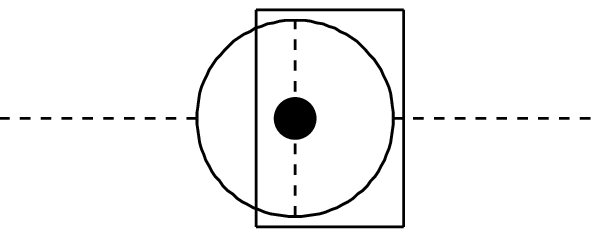} }  
    & 
    &  
  \\
    & $-$
    &       \raisebox{-.74cm}{ \epsfxsize = 3.3cm \epsffile{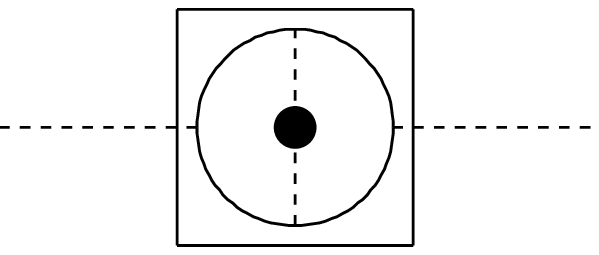} }  
    & $+ ~2 \times$ & \raisebox{-.74cm}{ \epsfxsize = 3.3cm \epsffile{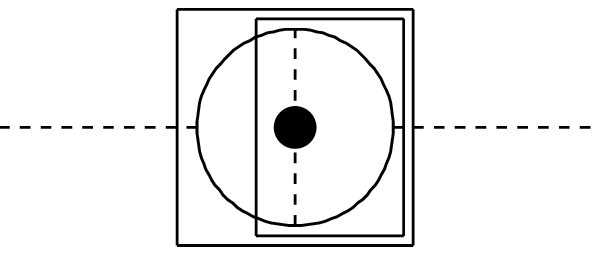} }  
    & $-$
    & overall subtraction
  \\
      $i \hat{A}_5(s) ~=$
    & 
    &       \raisebox{-.74cm}{ \epsfxsize = 3.3cm \epsffile{a5.eps} }  
    & $-$ & \raisebox{-.74cm}{ \epsfxsize = 3.3cm \epsffile{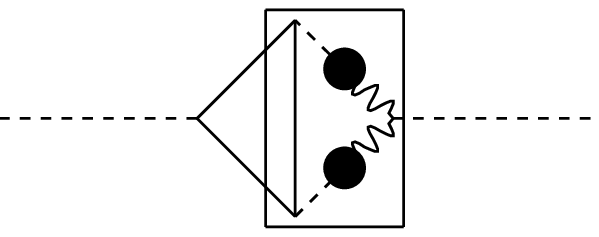} }  
    & 
    &  
  \\
    & $-$ & \raisebox{-.74cm}{ \epsfxsize = 3.3cm \epsffile{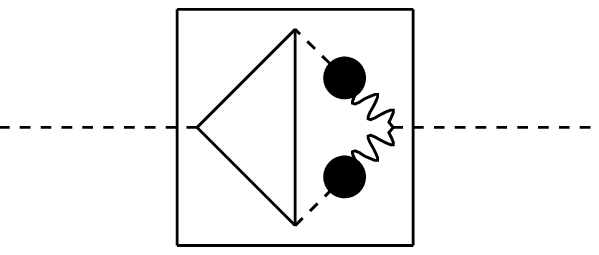} } 
    & $+$ & \raisebox{-.74cm}{ \epsfxsize = 3.3cm \epsffile{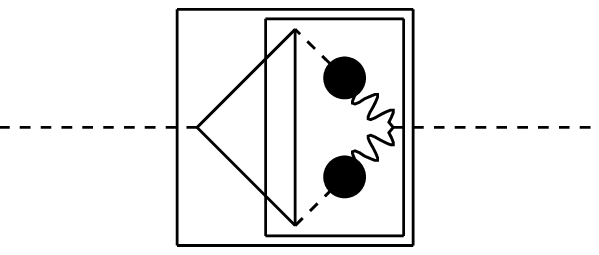} } 
    & 
    & 
\end{tabular}
\caption{{\em Definition of the subtracted $\chi \chi$ self-energy graphs.
              The symbol $\partial$ indicates the differentiation with
              respect to the external momentum of the box.}}
\end{figure}

\begin{figure}
\begin{tabular}{rcccccl}
      $i \frac{v^2}{N} \hat{B}_1(s) ~=$
    & 
    &       \raisebox{-.74cm}{ \epsfxsize = 3.3cm \epsffile{b1.eps} }  
    & $-$ & \raisebox{-.74cm}{ \epsfxsize = 3.3cm \epsffile{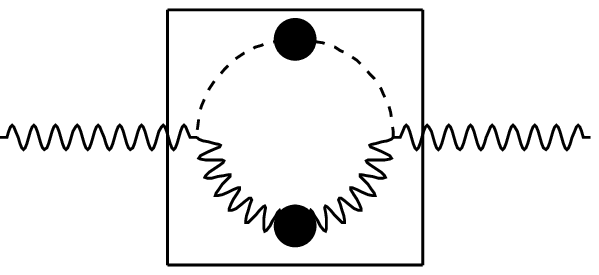} }  
    & $-$ & \raisebox{-.74cm}{ \epsfxsize = 3.3cm \epsffile{b1c.eps} \hspace{-1.9cm} \raisebox{.5cm}{$\partial$} } 
  \\
      $i \frac{v^2}{N} \hat{B}_2(s) ~=$
    & 
    &       \raisebox{-.74cm}{ \epsfxsize = 3.3cm \epsffile{b2.eps} }  
    & $-$ & \raisebox{-.74cm}{ \epsfxsize = 3.3cm \epsffile{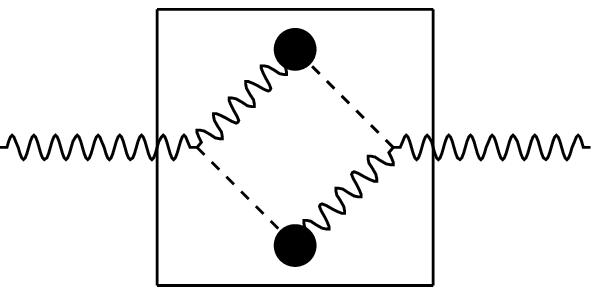} }  
    & 
    & 
  \\
      $i \frac{v}{\sqrt{N}} \hat{C}_1(s) ~=$
    & 
    &       \raisebox{-.74cm}{ \epsfxsize = 3.3cm \epsffile{c1.eps} }  
    & $-$ & \raisebox{-.74cm}{ \epsfxsize = 3.3cm \epsffile{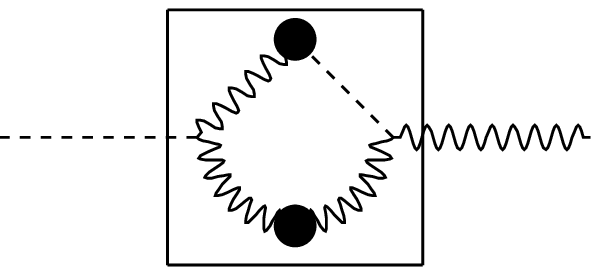} }  
    & 
    & 
  \\
      $i \frac{v}{\sqrt{N}} \hat{C}_2(s) ~=$
    & 
    &       \raisebox{-.74cm}{ \epsfxsize = 3.3cm \epsffile{c2.eps} }  
    & $-$ & \raisebox{-.74cm}{ \epsfxsize = 3.3cm \epsffile{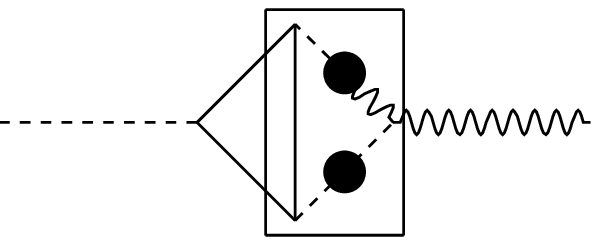} }  
    &
    & 
  \\
    & $-$ & \raisebox{-.74cm}{ \epsfxsize = 3.3cm \epsffile{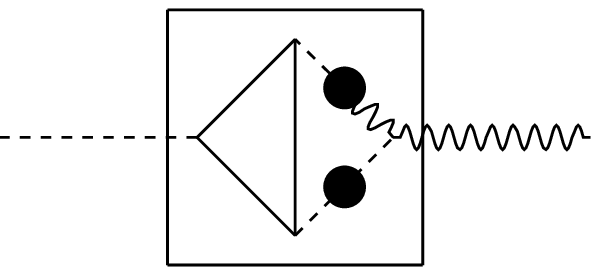} } 
    & $+$ & \raisebox{-.74cm}{ \epsfxsize = 3.3cm \epsffile{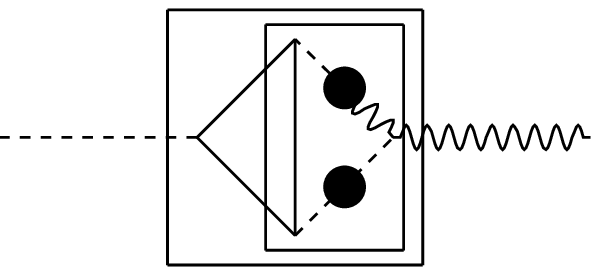} } 
    & 
  \\
      $i \frac{v^2}{N} \hat{D}(s) ~=$
    & 
    &       \raisebox{-.74cm}{ \epsfxsize = 3.3cm \epsffile{d.eps} }  
    & $-$ & \raisebox{-.74cm}{ \epsfxsize = 3.3cm \epsffile{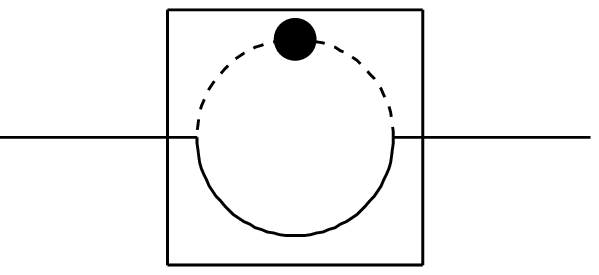} }  
    & $-$ & \raisebox{-.74cm}{ \epsfxsize = 3.3cm \epsffile{dc1.eps} \hspace{-1.9cm} \raisebox{.4cm}{$\partial$} } 
\end{tabular}
\caption{{\em Definition of the subtracted $\sigma \sigma$, 
              $\chi \sigma$ and $\pi \pi$ self-energy graphs.}}
\end{figure}

\begin{figure}[t]
\begin{tabular}{cccccccc}
      $i \frac{v}{\sqrt{N}} \hat{E}_1(s)$
    & $=$
    & 
    &       \raisebox{-.74cm}{ \epsfxsize = 3.3cm \epsffile{e1.eps} }  
    & $-$ & \raisebox{-.74cm}{ \epsfxsize = 3.3cm \epsffile{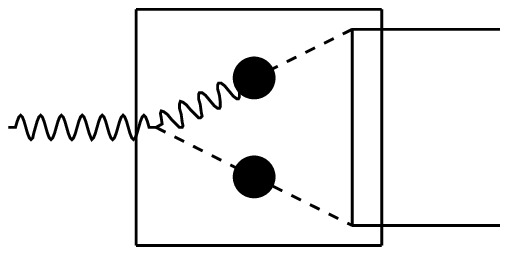} }  
    & 
    & 
  \\
      $i \frac{v}{\sqrt{N}} \hat{F}_1(s)$
    & $=$
    & 
    &       \raisebox{-.74cm}{ \epsfxsize = 3.3cm \epsffile{f1.eps} }  
    & $-$ & \raisebox{-.74cm}{ \epsfxsize = 3.3cm \epsffile{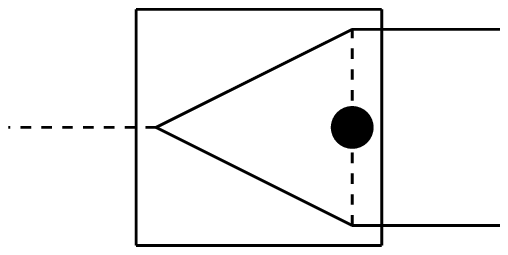} }  
    & 
    & 
  \\
      $i \frac{v}{\sqrt{N}} \hat{F}_2(s)$
    & $=$
    & 
    &       \raisebox{-.74cm}{ \epsfxsize = 3.3cm \epsffile{f2.eps} }  
    & $-$ & \raisebox{-.74cm}{ \epsfxsize = 3.3cm \epsffile{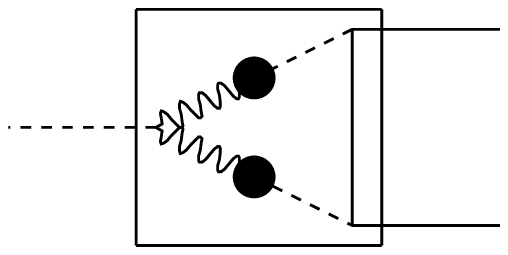} }  
    & 
    & 
  \\
      $i \frac{v}{\sqrt{N}} \hat{F}_3(s)$
    & $=$
    & 
    &       \raisebox{-.74cm}{ \epsfxsize = 3.3cm \epsffile{f3.eps} }  
    & $-$ & \raisebox{-.74cm}{ \epsfxsize = 3.3cm \epsffile{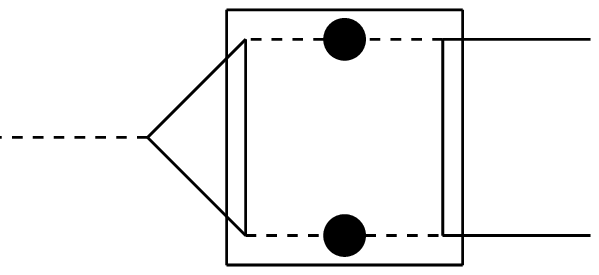} }  
    & 
    & 
  \\
    &       
    & $-$ & \raisebox{-.74cm}{ \epsfxsize = 3.3cm \epsffile{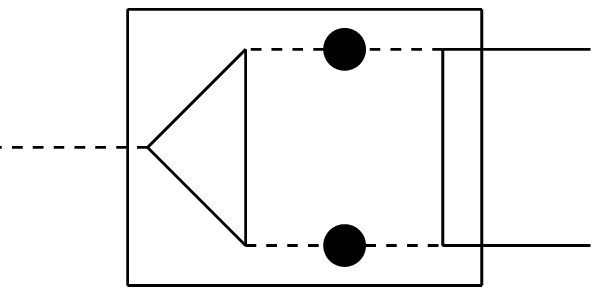} } 
    & $+$ & \raisebox{-.74cm}{ \epsfxsize = 3.3cm \epsffile{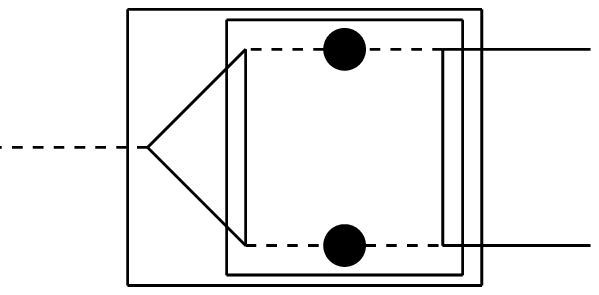} } 
    & 
    &  
\end{tabular}
\caption{{\em Definition of the subtracted vertex graphs.}}
\end{figure}

The numerical evaluation of the ultraviolet finite, subtracted graphs
is done by using a numerical method for the calculation of massive
three-loop Feynman diagrams \cite{3loop}. This method reduces all
subtracted graphs to a two-dimensional integral representation.
After an appropriate rotation of the integration path in the 
complex plane, these two-dimensional integrals can be evaluated
numerically \cite{oneovn}.

The subtracted graphs are used further for calculating physical
amplitudes. Here we consider the scattering processes 
$f\bar{f} \rightarrow H \rightarrow f^{\prime}\bar{f^{\prime}}$ and
$f\bar{f} \rightarrow H \rightarrow ZZ, WW$. Phenomenologically 
they are important as a Higgs production mechanism at a possible
muon collider. Also the heavy Higgs effects in these scattering
processes are related to those in the gluon fusion process,
which is the main Higgs production mechanism at the LHC.

To start with, we consider the matrix element 
for the fermion scattering process
$f\bar{f} \rightarrow H \rightarrow f^{\prime}\bar{f^{\prime}}$
at leading order in the fermion mass. Then, 
the correction to the $Hf\bar{f}$ vertex is given simply
by the ratio of the wave function renormalization factors
$Z_{\sigma}/Z_{\pi}$, because true vertex diagrams 
are of higher order in the fermion mass. The calculation
reduces essentially to evaluating the Higgs propagator.
Up to the overall factor from the tree level Yukawa couplings, the amplitude
is given by the following expression, were we included the
relevant counterterms in addition to the pure $1/N$ multiloop 
graphs of fig. 3: 

\begin{eqnarray}
{\cal M}_{f\bar{f}} & = &
\frac{1}{s - m^2(s) \left[ 1 - \frac{1}{N} f_1(s)  \right] }
  \nonumber  \\
f_1(s)  & = &
         \frac{m^2(s)}{v^2} \left[ \alpha(s) + \delta \lambda 
                                             + 2 \frac{v^2}{m^2(s)} \delta Z_{\chi} 
                                             + 8 K \alpha_0^2(s)
                                             + 2 \alpha_0(s) \delta Z_{\pi}
                            \right]
  \nonumber  \\
  & &
       + 2                  \left[ \gamma(s)- \delta Z_{\chi} - 8 K \alpha_0(s) 
                            \right]
       + \frac{v^2}{m^2(s)} \left[ \beta(s) - \delta t_{\chi} + 8 K + 2 \frac{s}{v^2} \delta Z_{\pi}
                            \right]
\end{eqnarray}

One can easily see in the above expression that one could have done 
without a wave function renormalization for the unphysical $\chi$
field, and also without the $K$ counterterm which was introduced
in eq. 6 only for convenience purposes. It can be seen easily 
that these terms cancel out trivially in the expression above.
In fact, since there is no Higgs external leg in the
process considered, it is also unnecessary to introduce a 
wave function renormalization for the $\sigma$ field.

As we discussed in section 3.1, these counterterms are
only introduced for convenience. If one cancels out these
spurious terms in the expression above, one is left with
a sum of multiloop diagrams which are individually ultraviolet
divergent. Since the whole expression is a physical quantity,
the ultraviolet divergencies cancel out among the multiloop 
diagrams. However, the actual cancellation pattern is not
very transparent due to the complexity of the diagrams.
Because we need to calculate the $1/N$ graphs numerically,
we need the expressions to be ultraviolet convergent.
At the same time it is very complicated to extract the
ultraviolet divergencies and subdivergencies from the graphs 
of fig. 3 as $1/\epsilon$ poles, as it is done in usual
Feynman diagrams.
In more complicated processes, such as three- and four-point
processes, the cancellation is even more involved.

The $K$ and $\delta Z_{\chi}$ counterterms serve as vehicles
of the ultraviolet cancellations among the multiloop diagrams.
For this purpose we assign to these 
two counterterms the following expressions:

\begin{eqnarray}
    K              & = &  - \frac{1}{4} B_2(0)  \nonumber \\
  \delta Z_{\chi}  & = & C_1(0) + C_2(0) + 2 B_2(0) \alpha_0(\mu^2 \rightarrow 0)  ~~~~~~,
\end{eqnarray}
and also we note the identity:

\begin{equation}
  \delta t_{\chi} = B_1(0) - B_2(0)  ~~~~~~.
\end{equation}

Then the actual multiloop $1/N$ graphs from $\alpha$, $\beta$ and $\gamma$
combine with the counterterms and give precisely the subtracted multiloop
graphs $\hat{A}_i$, $\hat{B}_i$ and $\hat{C}_i$ defined in figures 5 and 6:

\begin{equation}
f_1(s)   = 
         \frac{m^2(s)}{v^2} \left[ \hat{\alpha}(s) + \delta \lambda^{fin.}
                            \right]
       + 2                 \hat{\gamma}(s)
       + \frac{v^2}{m^2(s)} \left[ \hat{\beta}(s)
                               - 2 \frac{s-m^2(s)}{v^2} \left( \delta Z_{\sigma} - \delta Z_{\pi} \right)
                            \right]
\end{equation}
Actually the finite term $\left( \delta Z_{\sigma} - \delta Z_{\pi} \right)$ 
in this expression simply means that one has to subtract the momentum
derivative of diagram $D(s)$ from diagram $B_1(s)$, rather than the derivative
of $B_1(s)$, as defined in figure 6. The finite contribution 
$\delta \lambda^{fin.}$
which is left in eq. 15 simply reminds that for specifying the strength
of the coupling of the theory, a mass scale needs to be given along with 
the value of $\lambda$. A shift in $\delta \lambda^{fin.}$ can be 
absorbed into a shift in the subtraction point $\mu$. As such,
$\delta \lambda^{fin.}$ can be shifted to zero. 

Along the same lines, the following expression is 
obtained for the amplitude of the 
$f\bar{f} \rightarrow H \rightarrow ZZ, WW$
scattering process:

\begin{eqnarray}
{\cal M}_{WW} & = &
\frac{m^2(s)}{\sqrt{N} v}
\frac{1 - \frac{1}{N} f_2 }{s - m^2(s) \left[ 1 - \frac{1}{N} f_1(s)  \right] }
  \nonumber  \\
f_1(s)  & = &
           \frac{m^2(s)}{v^2} \, \hat{\alpha}(s)
       + 2 \hat{\gamma}(s)
       +   \frac{v^2}{m^2(s)} \left[ \hat{\beta}(s)
                               - 2 \frac{s-m^2(s)}{v^2} \left( \delta Z_{\sigma} - \delta Z_{\pi} \right)
                              \right]
  \nonumber  \\
f_2(s)  & = &
           \frac{m^2(s)}{v^2} \, \hat{\alpha}(s) 
       +   \hat{\gamma}(s)
       -   \hat{\phi}(s)
       -   \frac{v^2}{m^2(s)} \hat{\eta}(s)
\end{eqnarray}

All quantities involved in this expression, 
$\hat{\alpha}$, $\hat{\beta}$, $\hat{\gamma}$, $\hat{\eta}$, and $\hat{\phi}$, 
are sums of the subtracted graphs defined in figures 4, 5 and 6. 
They can be calculated directly by 
numerical integration \cite{oneovn,thomas:rev}.

\subsection{The saturation effect}

The shape of the Higgs resonance can be obtained nonperturbatively 
in the quartic Higgs coupling at next-to-leading order in the 
$1/N$ expansion by evaluating numerically the expressions for
${\cal M}_{f\bar{f}}$ and ${\cal M}_{WW}$ given in the previous section.
These two scattering processes are the main production and decay modes 
for the Higgs boson at a possible muon collider. Also these processes
are related to Higgs production by gluon fusion \cite{gluefusion}. 
This is the main 
production mechanism at hadron colliders such as the LHC.

We give in figure 8 the resulting line shapes of the Higgs resonance.
One feature of these line shapes is that they agree remarkably well
with the perturbative results for low couplings. The next-to-leading
order $1/N$ results are very close to the two-loop perturbation theory
line shapes for $m_H$ up to about 800--900 GeV. The agreement confirms
the consistency of the $1/N$ approach in higher orders and establishes that
the next-to-leading oder is an excellent approximation.

\begin{figure}
\hspace{1.5cm}
    \epsfxsize = 15cm
    \epsffile{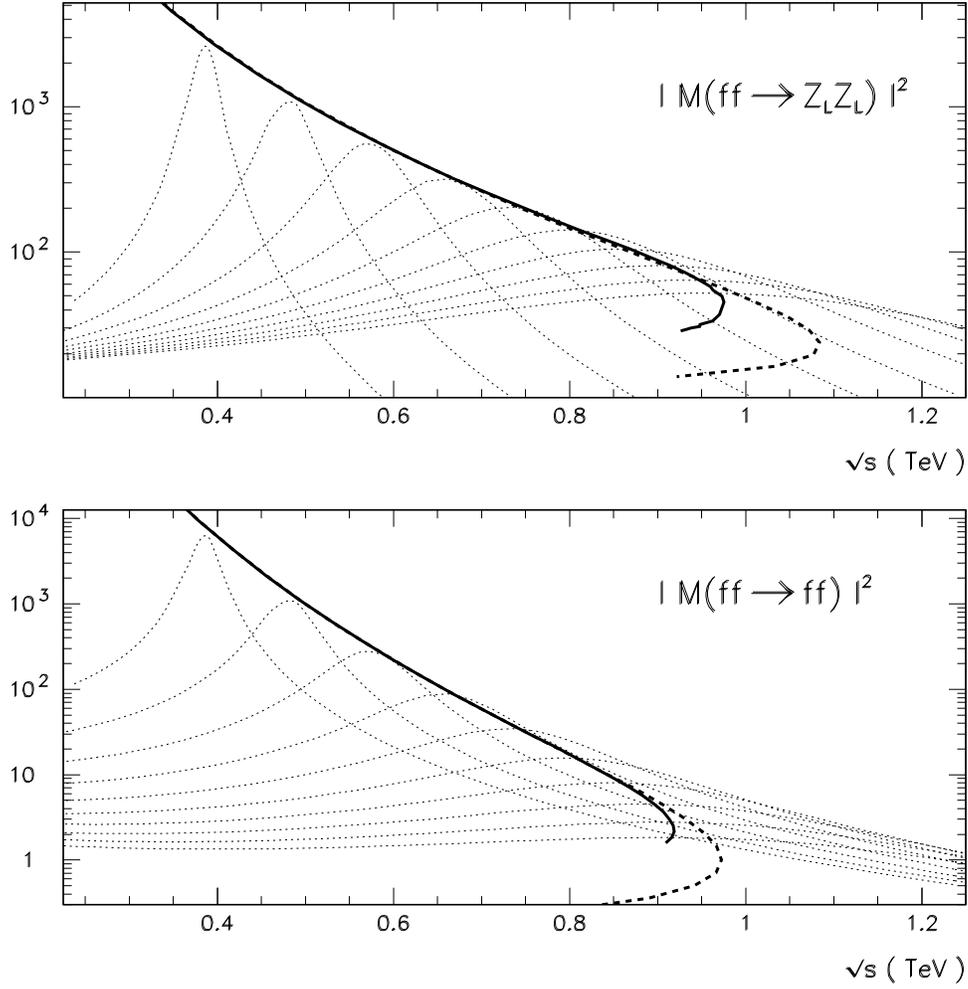}
\caption{{\em The line shape of the Higgs resonance in the scattering
              processes
              $f\bar{f} \rightarrow H \rightarrow f^{\prime}\bar{f^{\prime}}$ 
              and
              $f\bar{f} \rightarrow H \rightarrow ZZ, WW$.
              We marked the position of the maxima of the resonances
              (solid line for the $1/N$ result and dashed line for the
               perturbative result at two-loop).}}
\end{figure}

At higher quartic coupling a saturation effect sets in. The maximum
of the resonance does not shift towards higher energy, the resonance 
only becomes wider. To make this effect clearer, we extract some
mass ($M_{peak}$) and width ($\Gamma_{peak}$) 
variables from the position and the height of the
resonance in fermion scattering. The definition of the
variables $M_{peak}$ and $\Gamma_{peak}$ is the following.
We determined numerically the position and the height of
the maximum of the resonance. Then, $M_{peak}$ and $\Gamma_{peak}$ 
are the mass and width of a Breit-Wigner resonance which
has the same height and position of the peak.
Of course, the actual Higgs lineshapes are not exactly
of Breit-Wigner type. However, $M_{peak}$ and $\Gamma_{peak}$ 
describe reasonably well the main features of the resonance.
We compare in figure 9 
the $M_{peak}$ -- $\Gamma_{peak}$ relation with the perturbative result.
Of course one is free to choose any other parameterization of the resonance,
but the $M_{peak}$ and $\Gamma_{peak}$
variables which we use here are sufficient for comparing
with perturbation theory.

\begin{figure}
\hspace{1.5cm}
    \epsfxsize = 15cm
    \epsffile{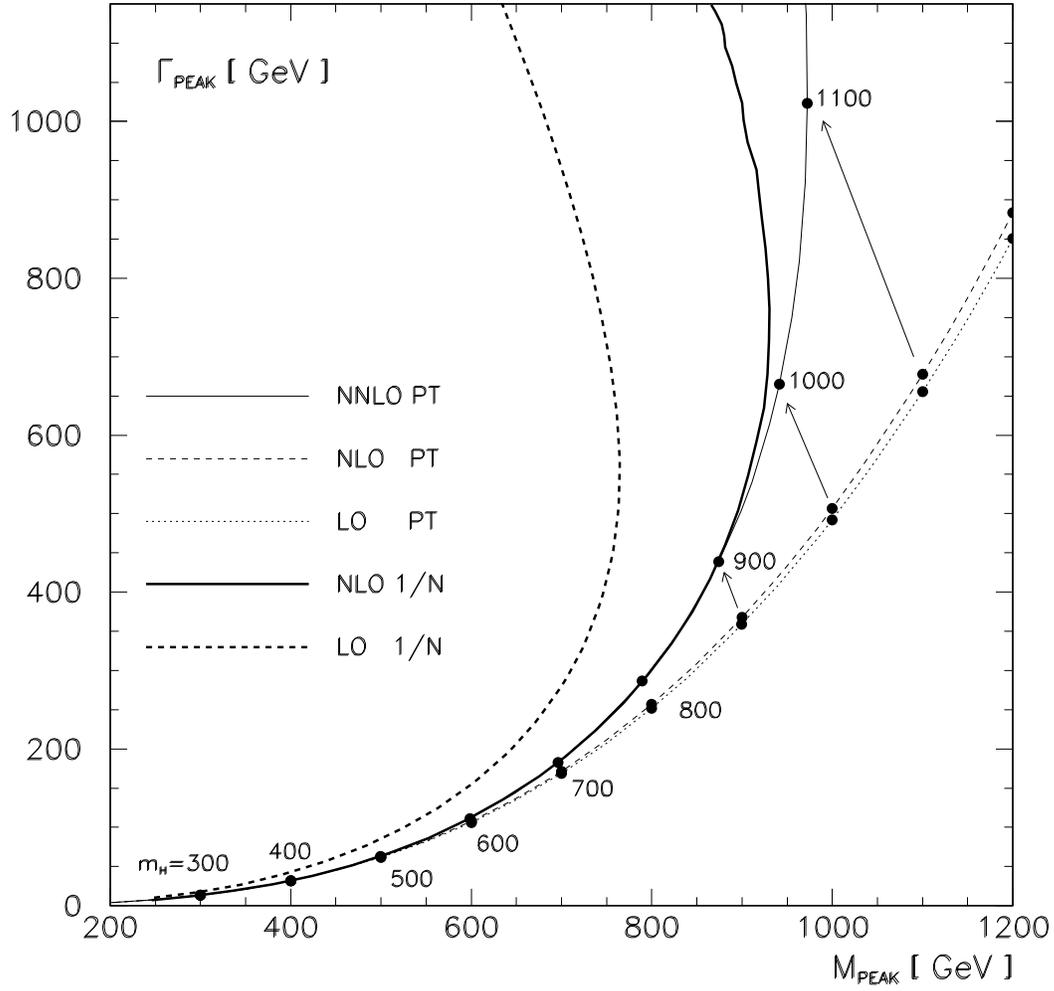}
\caption{{\em The saturation effect in the
              $f\bar{f} \rightarrow H \rightarrow f^{\prime}\bar{f^{\prime}}$ 
              scattering. The mass and width variables  
              $M_{peak}$ and $\Gamma_{peak}$ are related to the
              position and the height of the Higgs resonance as
              explained in the text.}}
\end{figure}

As one can see in figure 8, the saturation effect is present in the
$f\bar{f} \rightarrow H \rightarrow ZZ, WW$ 
scattering process as well, at a comparable energy. The precise
maximum position of the peak is process dependent because the
resonance shape is deformed by the energy dependence of different
contributions, such as the vertex corrections in this case. 
A universal way of parameterizing the saturation effect would
be by using the pole mass and width of the Higgs particle.
Extracting this in the next-to-leading $1/N$ approach is
numerically more difficult because the subtracted $1/N$ graphs
need to be continued into the second Riemann sheet for solving 
the pole equation. This is not available yet.

The saturation effect provides more insight in the way perturbation
theory breaks down and radiative corrections blow up in the Higgs sector.
The dramatical failure of perturbation theory at around 1 TeV 
is well-established
by tree level unitarity violations \cite{lee-quigg}. 
Higher order radiative corrections
blow up at a similar scale, as was discussed in section 2. 
Less violent problems can show up in the form of considerable scheme
ambiguities of the results. At the same time however, it
is puzzling that at the 1 TeV scale the quartic coupling is numerically
not exceedingly large yet. The quartic coupling only becomes of the
order of unity when the tree level on-shell Higgs mass is of the
order of 1.5 TeV. Naively this is where one would expect heavy problems
for perturbation theory to set in. From the perspective of the saturation effect,
the explanation is clear now. The mass of the Higgs boson is not a good
expansion parameter beyond the saturation point. In the saturation region
the mass does not increase when the coupling is enhanced. In fact,
the width of the resonance as a function of the mass 
is a double valued function.
The well-known perturbation theory results break down at about 1 TeV
because of the use of a bad expansion parameter. For that region the width
would be a more appropriate parameterization of the coupling than the mass.


\section{Nonperturbative solutions of other theories: $1/N$ QED}

In this section we would like to comment briefly on the perspectives 
of extending this kind of nonperturbative methods to solving other
field theories of physical interest.

We already mentioned that the applicability of the nonperturbative
$1/N$ expansion, especially in higher orders, depends crucially on the particular
theory under consideration. One needs to arrange the perturbative
expansion of the theory in such a way that the various loop
contributions can be sorted out by powers of $1/N$ in a manageable way,
so that the graphs of all loop-orders can be explicitly calculated and 
summed up for a given order of $1/N$.

Such a theory where the rearrangement of perturbation theory is
straightforward is the ordinary QED. QED can be seen as
an example where the $1/N$ expansion works poorly because the
value of $N$ is too small. The existing
calculations in perturbative QED are rather advanced, the 
coupling constant is small, and therefore the $1/N$ treatment 
of QED can hardly compete
with the perturbative treatment. Here we will briefly discuss a 
simple leading order $1/N$ QED calculation for comparing with perturbation
theory, and see in which cases a $1/N$ treatment can be superior
to perturbation theory.

QED can be organized as a $1/N$ expansion by introducing
$N$ species of electrons and at the same time dividing the gauge
coupling by $\sqrt{N}$. Then one sees immediately that the counting of
powers of $1/N$ proceeds similarly as in the case of the sigma model.
Ordinary QED is recovered in the limit $N=1$. Already at this point
one can expect the convergence of the $1/N$ expansion to be poor
because of the value of the expansion parameter.

\begin{figure}
 \hspace{3cm}
  \epsfxsize = 7cm \epsffile{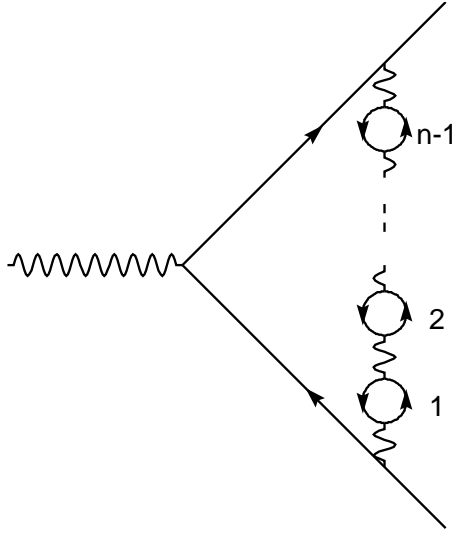}
\caption{{\em Diagrams which contribute to the anomalous
              magnetic moment of the electron in leading order
              of the $1/N$ expansion.}}
\end{figure}

We show in figure 10 the Feynman diagrams which contribute to the
anomalous magnetic moment of the electron in leading order of $1/N$.
They are the same diagrams discussed in ref. \cite{lautrup} in the context
of the large order behaviour of QED. We refer to this work for details
on the calculation of the contributions to the anomalous magnetic
moment due to these diagrams. The equivalent of the tachyonic regularization
in the $O(N)$-symmetric sigma model is the subtraction in the 
loop momentum integration of the diagrams of figure 10 of a term
of the type $e^{1/\alpha}$, which vanishes in perturbation theory,
as discussed in ref. \cite{lautrup}. Note that due to the value of the 
coupling constant, in QED the Landau pole is at a very high energy scale.

The perturbative results for the anomalous 
magnetic moment of the electron up to four-loop are \cite{remiddi}:

\begin{eqnarray}
a_{1-loop} &=& .5  \nonumber \\    
a_{2-loop} &=& -.328478965  \nonumber \\    
a_{3-loop} &=& 1.181241  \nonumber \\    
a_{4-loop} &=& -1.557     ~~~~,
\end{eqnarray}
and their contribution to the anomalous
magnetic moment is $\Sigma_n \, a_n (\alpha/\pi)^n$.

By numerical integration one can calculate the LO contribution in the
$1/N$ expansion to the anomalous magnetic moment of the electron.
The result is $a^{LO ~ 1/N} = .001161494$.
The first few terms in the loop expansion of this result are:

\begin{eqnarray}
a^{LO ~ 1/N}_{1-loop}    &=& .5           \nonumber \\    
a^{LO ~ 1/N}_{2-loop}    &=& .015687421   \nonumber \\    
a^{LO ~ 1/N}_{3-loop}    &=& .002558524   \nonumber \\    
a^{LO ~ 1/N}_{4-loop}    &=& .000876865   \nonumber \\    
a^{LO ~ 1/N}_{5-loop}    &=& .000470905   \nonumber \\    
a^{LO ~ 1/N}_{6-loop}    &=& .000344687   \nonumber \\    
a^{LO ~ 1/N}_{7-loop}    &=& .000318119   \nonumber \\    
a^{LO ~ 1/N}_{8-loop}    &=& .000352804   \nonumber \\    
a^{LO ~ 1/N}_{9-loop}    &=& .000455370   \nonumber \\    
a^{LO ~ 1/N}_{10-loop} &=& .000668824   \nonumber \\    
a^{LO ~ 1/N}_{11-loop} &=& .001099482   \nonumber \\    
a^{LO ~ 1/N}_{12-loop} &=& .001997590   \nonumber \\    
a^{LO ~ 1/N}_{13-loop} &=& .003971440    ~~~~,
\end{eqnarray}
such that their contribution to the anomalous
magnetic moment is $\Sigma_n \, a^{LO ~ 1/N}_n (\alpha/\pi)^n$.
The numbers above agree with the results given in ref. \cite{lautrup}.

By comparing the LO $1/N$ result with perturbation theory, one sees that
this is only marginally different from the one-loop perturbative result.
Comparing the  perturbative expansion of the LO $1/N$ result (eqns. 18) and
the perturbation theory result (eqns. 17), it can be seen that the LO $1/N$
Feynman diagrams are not numerically the main contribution from
higher-loop orders in perturbation theory.

This shows that the usefulness of the $1/N$ expansion 
versus perturbation theory 
depends crucially on the value of $N$ and of the coupling constant.
If $N$ is large enough, the other diagrams appearing in higher-order
perturbation theory besides those included in the $1/N$ expansion 
will be suppressed by powers of $1/N$. This way the $1/N$ expansion
can converge faster than perturbation theory. If the coupling constant is
large, perturbation theory starts to diverge already at low orders
and the scheme ambiguities may prevent one to obtain the desired
accuracy. Then the $1/N$ expansion is expected to be a better alternative
because it is free of scheme ambiguities and the solution is valid
for strong coupling as well. 

Therefore it is not surprising that the $1/N$ expansion works so well 
in heavy Higgs physics, where $N = 4$ and $\lambda$ is of order 1.

Extending these computational tools in the case of nonabelian
gauge theories is very difficult because of the presence of trilinear
and quartic couplings of the gauge bosons. In the case of QCD the
topological structure of the graphs which appear in the $1/N_c$ 
expansion is known \cite{thooft:1}, 
but so far they could not be calculated explicitly in four dimensions.
The well-known solution of QCD in two dimensions \cite{thooft:2} 
was possible because in this case the quartic and trilinear couplings
are absent, which reduces the types of topologies of Feynman graphs
appearing in higher loop orders.


\section{Conclusions}

We investigated in some detail the scalar sector of the standard model
at strong coupling. For doing this we used both perturbation
theory up to two-loop order and a nonperturbative treatment
within the next-to-leading order $1/N$ expansion. With these
approaches we treated the main heavy Higgs decay modes as well as
two scattering processes where the Higgs line shape can be observed, 
$f\bar f \rightarrow H \rightarrow f^{\prime} \bar{f^{\prime}}$ and 
$f\bar f \rightarrow H \rightarrow ZZ, WW$. These two scattering processes
are the main $s$ channel production modes of the Higgs boson at
muon colliders, and can also be related to the heavy Higgs effects
in the gluon fusion process, which is the dominant production
mechanism at the LHC. 

The results show in all cases a very good agreement between
perturbation theory and the $1/N$ expansion up to 800--900 GeV.
They confirm the existence of a Higgs mass saturation effect
in both scattering processes analyzed, which seems to be a
general feature of resonant Higgs processes. The nonperturbative 
mass saturation value is just under 1 TeV for both processes.
As a preliminary study based on perturbation theory has 
shown, it is expected that the nonperturbative
mass saturation effect will play an important role in
the experimental strategy for heavy Higgs searches at the LHC.

A comparison of the nonperturbative behaviour at strong coupling
to the pattern observed in perturbation theory in higher 
loop orders suggests that radiative corrections in the scalar
sector blow up very strongly at about 1 TeV mainly because of 
the nonperturbative mass saturation effect, rather than
because of a genuinely strong coupling. Due to this dynamical
effect the mass of the Higgs boson, although widely used
as a parameterization of the coupling in phenomenological 
studies so far, is not an appropriate parameter in the
saturation region.

On the theoretical side, we have shown that in the case 
of the $O(N)$-symmetric sigma model, a reliable nonperturbative
solution can be obtained by using a higher order $1/N$ expansion,
which is free of renormalization scheme ambiguities and which
is valid at strong coupling as well. We discussed the tachyonic 
regularization which we introduced for calculating higher orders
in the $1/N$ expansion. We also discussed the renormalization
within the $1/N$ expansion in higher orders.

Finally, we discussed briefly the applicability of such
nonperturbative methods to other theories of physical
interest. 


\vspace{.5cm}

{\bf Acknowledgement}

A.G. is grateful to Bernd Kniehl and Alberto Sirlin for insightful remarks.



\end{document}